\documentclass[twocolumn,tighten,times]{aastex62}

\newcommand{\msol}{\mbox{$M_\odot$}}

\newcommand{\HII}{H {\small{II}} }
\newcommand{\HI}{H\,{\sc i} }
\newcommand{\kms}{\rm km~s^{-1}}
\newcommand{\Msun} {M_{\sun}}

\begin{document}
\title{Discovery of Extra-Planar \HI Clouds and a \HI Tail in the M101 Galaxy Group with FAST }

\correspondingauthor{Jin-Long Xu}
\email{xujl@bao.ac.cn}

\author{Jin-Long Xu}
\affiliation{National Astronomical Observatories, Chinese Academy of Sciences, Beijing 100101, People's Republic of China}
\affil{CAS Key Laboratory of FAST, National Astronomical Observatories, Chinese Academy of Sciences, Beijing 100101, People's Republic of China}

\author{Chuan-Peng Zhang}
\affiliation{National Astronomical Observatories, Chinese Academy of Sciences, Beijing 100101, People's Republic of China}
\affil{CAS Key Laboratory of FAST, National Astronomical Observatories, Chinese Academy of Sciences, Beijing 100101, People's Republic of China}

\author{Naiping Yu}
\affiliation{National Astronomical Observatories, Chinese Academy of Sciences, Beijing 100101, People's Republic of China}
\affil{CAS Key Laboratory of FAST, National Astronomical Observatories, Chinese Academy of Sciences, Beijing 100101, People's Republic of China}

\author{Ming Zhu}
\affiliation{National Astronomical Observatories, Chinese Academy of Sciences, Beijing 100101, People's Republic of China}
\affil{CAS Key Laboratory of FAST, National Astronomical Observatories, Chinese Academy of Sciences, Beijing 100101, People's Republic of China}

\author{Peng Jiang}
\affiliation{National Astronomical Observatories, Chinese Academy of Sciences, Beijing 100101, People's Republic of China}
\affil{CAS Key Laboratory of FAST, National Astronomical Observatories, Chinese Academy of Sciences, Beijing 100101, People's Republic of China}

\author{Jie Wang}
\affiliation{National Astronomical Observatories, Chinese Academy of Sciences, Beijing 100101, People's Republic of China}

\author{Xin Guan}
\affiliation{National Astronomical Observatories, Chinese Academy of Sciences, Beijing 100101, People's Republic of China}
\affil{CAS Key Laboratory of FAST, National Astronomical Observatories, Chinese Academy of Sciences, Beijing 100101, People's Republic of China}

\author{Xiao-Lan Liu}
\affiliation{National Astronomical Observatories, Chinese Academy of Sciences, Beijing 100101, People's Republic of China}
\affil{CAS Key Laboratory of FAST, National Astronomical Observatories, Chinese Academy of Sciences, Beijing 100101, People's Republic of China}

\author{Xiaolian Liang}
\affiliation{National Astronomical Observatories, Chinese Academy of Sciences, Beijing 100101, People's Republic of China}
\affil{University of Chinese Academy of Sciences, Beijing 100049, People's Republic of China}

\author{the FAST Collaboration}
\affiliation{National Astronomical Observatories, Chinese Academy of Sciences, Beijing 100101, People's Republic of China}

\begin{abstract}
We present a new high-sensitivity \HI observation toward nearby spiral galaxy M101 and its adjacent  2$^{\circ}\times$ 2$^{\circ}$ region using the Five-hundred-meter Aperture Spherical radio Telescope (FAST). From the observation, we detect a more extended and asymmetric \HI disk around M101. While the  \HI velocity field within the M101's optical disk region is regular, indicating that the relatively strong disturbance occurs in its outer disk. Moreover, we identify three new \HI clouds located on the southern edge of the M101's \HI disk. The masses of the three \HI clouds are  1.3$\times$10$^{7}$ \msol, 2.4$\times$10$^{7}$ \msol, and 2.0$\times$10$^{7}$ \msol, respectively. The \HI clouds similar to dwarf companion NGC 5477 rotate with the \HI disk of M101. Unlike the NGC 5477, they have no optical counterparts.  Furthermore, we detect a new \HI tail in the extended \HI disk of M101. The \HI tail detected gives a reliable evidence for M101 interaction with the dwarf companion NGC 5474. We argue that the extra-planar gas (three \HI clouds) and the \HI tail detected in the M101's disk may  origin from a minor interaction with NGC 5474. 
 
\end{abstract}


\section{Introduction} \label{sec:intro}
Galaxies show an almost constant star formation rate throughout the Hubble time \citep{Panter2007}, such as the Milky Way. In order to maintain the stable star formation and prevent gas from exhaustion, these galaxies need a continuous supply of fresh gas. Theoretical arguments predict that galaxies can continually accrete gas from their surroundings, as material falls in from the more diffuse intragroup medium \citep{Larson1972, Sommer-Larsen2006, Mihos2012}. However, direct observation of the actual accretion processes is very difficult \citep{Sancisi2008,Pezzulli2016}. \citet{Sancisi2008} pointed out that  interaction, minor merger, and peculiar \HI structure around galaxies can be considered as the ongoing  or recent precesses of the direct accretion. In the galaxies, the neutral hydrogen (\HI) emission  is more extended than their optical counterparts. Deep \HI observations can reveal faint gaseous distribution around the galaxies, hence it can provide an important tool for uncovering the accretion process. Especially, nearby galaxies are unique laboratories for examining the processes of galaxy accretion. 

Nearby spiral galaxy M101 (NGC 5457) is the nearest massive ScI galaxy \citep{Huchtmeier1979}. The distance to M101 is 6.9 Mpc \citep{Matheson2012}. M101 is the dominant member of a small group of galaxies.  In recent years M101 has been well studied in detail at many wavelengths. The most striking feature that M101 shows a distorted \HI disk \citep{Walter2008}, and an asymmetric outer \HI plume to the southwest \citep{Huchtmeier1979}, indicating that M101 is interacting with other galaxies within the group \citep{Walter2008}. The two closest companions to M101 are NGC 5477 and NGC 5474. NGC 5477 resides near the northwestern spiral arm of M101, while  NGC 5474 with a strongly asymmetric disk \citep{Rownd1994,Kornreich1998} is located to the south \citep{Mihos2012}. Because  NGC 5477 has a low luminosity ($M_{V}$ =-15.3),  it can't be massive enough to creat such a strong morphological response in a giant Sc spiral like M101 \citep{Mihos2012}. Although the diffuse \HI  gases are found between M101 and NGC 5474 \citep{Huchtmeier1979,Mihos2012}, they were still unable to detect a characteristics of typical galactic interactions. 

In this paper, to explore the accretion process happened on M101,  we performed a new high-sensitivity \HI observation toward this galaxy and its adjacent region with FAST. 

\section{Observation and data processing}
\subsection{Observation}
To show the \HI (1420.4058 MHz) gas distribution surrounding M101, we mapped a  2$^{\circ}\times$ 2$^{\circ}$  region centered at position of R.A.= 14:03:12.6 and Dec.= 54:08:34.4 using the Five-hundred-meter Aperture Spherical radio Telescope (FAST), during  January and April  2021. FAST is located in Guizhou, China. The aperture of the telescope is 500 m, and its effective aperture is about 300 m. The 19-beam array receiver system in dual polarization mode is used as front end. It formally works in the frequency range from 1050 MHz to 1450 MHz. For spectral-line observation, FAST is equipped with two digital backend systems, Spec(F) and Spec(W+N). We select Spec(F) backend, which has a bandwidth of 500 MHz and 1048576 channels, resulting a frequency resolution of 476 Hz and corresponding to a  velocity resolution 0.1 km s$^{-1}$ at 1.4 GHz. 

Mapping observation uses the Multibeam on-the-fly (OTF) mode. This mode is proposed to map the sky with 19 beams simultaneously, and  has a similar scanning trajectory. The half-power beam width (HPBW) is about 2.9$^{\prime}$ at 1.4 GHz for each beam.  The pointing accuracy of the telescope was better than 8$^{\prime\prime}$.  To satisfy Nyquist sampling, the parameter of rotation angle is available in this mode. If the scan is along the latitude lines, the receiver platform needs to turn 23.4$^{\circ}$. While along the longitude lines, it needs to turn 53.4$^{\circ}$. After turning the angle, an interval between two adjacent parallel scans is about 1.14$^{\prime}$. Besides, we set the scan velocity of 15$^{\prime\prime}$ s$^{-1}$ and an integration time of 1 s per spectrum. During observations, system temperature was around 25 K. For intensity calibration, noises signal with amplitude of 10 K was injected under a period of two seconds. A total of 67 minutes is needed to acquire a 2$^{\circ}\times$ 2$^{\circ}$ map. In order to improve sensitivity, we observed two times for M101 with the Multibeam OTF mode. \citet{Jiang2019,Jiang2020} gave more details of the FAST instrument. 

\subsection{Data reduction}
For each spectrum with a bandwidth of 500 MHz, we intercept 2.2 MHz data for processing, which can cover the whole velocity range of M101. The data reduction was performed with the PYTHON package and GILDAS\footnote{http://www.iram.fr/IRAMFR/GILDAS} package. First of all, we will calibrate the unit of the spectrum into kelvin for each polarization signal through the following transformation,
\begin{equation}\label{eq:4}
      T_{\rm A} = T_{\rm cal}\frac{P_{\rm off}}{P_{\rm on}-P_{\rm off}},
\end{equation}
where $T_{A}$ is calibrated antenna temperature, while $T_{\rm cal}$ is noise diode temperature. $P_{\rm on}$ and $P_{\rm off}$ are power values when noise diode is on and off, respectively. The correction for the line intensity to brightness temperature ($T_{B}$) was made using the formula $T_{\rm B}= T_{\rm A}/\eta_{_{\rm MB}}$. Here $\eta_{_{\rm MB}}$ is the main beam efficiency, which can be determined by  $\eta_{_{\rm MB}}(\lambda)=0.8899[\theta_{_{\rm MB}}/(\lambda/D)]^{2}\eta_{_{\rm A}}$ \citep{Downes1989}, where $\theta_{_{\rm MB}}$ is beamwidth in radians, $D$ is the effective aperture, $\lambda$ is the observing wavelength, and $\eta_{_{\rm A}}$ is the aperture efficiency. The $\theta_{_{\rm MB}}$ and  $\eta_{_{\rm A}}$ values for each beam can be derived from \citet{Jiang2020}. Here we only give  the main beam efficiency of the central beam (Beam 1), $\eta_{_{\rm MB}}$ is $\sim$ 0.75 at 1.4 GHz.

Using the ArPLS algorithm, we mitigate radio frequency interference (RFI) by a fitting procedure of the data in the time-frequency domain. After combining the two polarizations, a first-order polynomial was used to fit baseline structure and subtract continuum sources to a line-free range of each spectrum. The L-band 19-beam receiver of FAST has a low level ($<$5\%) of sidelobes near the frequency of 1420 MHz \citep{Jiang2020,Han2021}, hence we  don't make the sidelobe correction.  Once the spectra have been fully calibrated, we apply them to a grid in the image plane. Each grid is adopted as 1.5$^{\prime}$.  This grid can be made in a TAN projection projection in the World Coordinate System as described by \citet{Calabretta2002}. 

Finally,  we make the calibrated data into the standard fits-cube format. For further processing of fits-cube data, we use GILDAS package, such as channel merging and regriding with a pixel of $1.5^{\prime}\times1.5^{\prime}$. The velocity resolution is smoothed to 2.1 $\kms$, then the noise RMS is about 56 mk (2.6 mJy beam$^{-1}$) for the M101 region. To facilitate comparison with the previous observations of M101 by the HI4PI data \citep{Collaboration2016}, the  FAST cube data are smoothed and resampled to the same angular resolution and pixel scale with the HI4PI data. Figure 1 shows comparison of the \HI spectrum of NGC 5474 in the M101 galaxy group from the FAST map (black line) and HI4PI map (red line). Both spectra are integrated over whole the NGC 5474's region. It is  clear that the two spectra agree within the noise.

\begin{figure}
\centering
\includegraphics[width = 0.45 \textwidth]{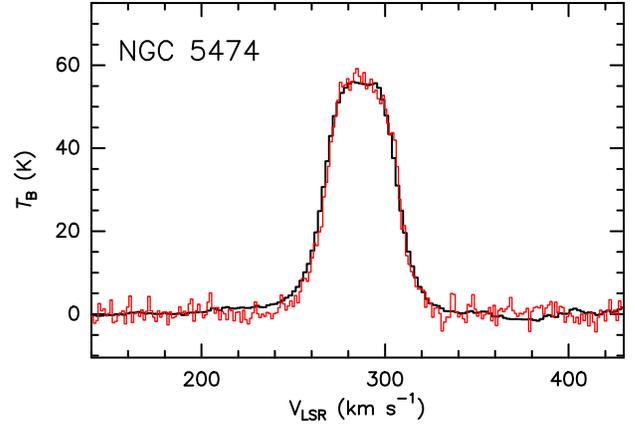}
\vspace{-2mm}
\caption{Comparison of the \HI spectrum of NGC 5474 in the M101 galaxy group from the FAST map (black line) and HI4PI map (red line). Both spetra are integrated over the NGC 5474's region.}
\label{Fig:Spectrum}
\end{figure}

\begin{figure}
\centering
\includegraphics[width = 0.48 \textwidth]{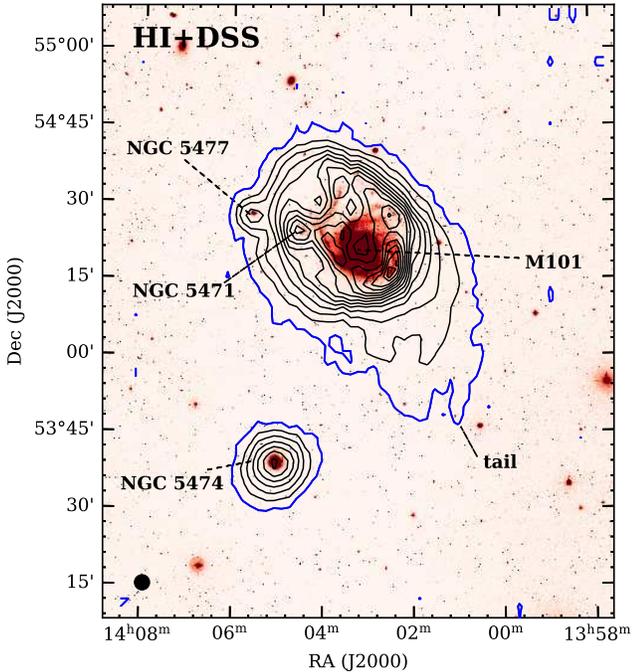}
\vspace{-7mm}
\caption{\HI column-density contours in blue and black colours, overlaid on the DSS  $B$-band optical image of the M101 galaxy group. The blue contour is 7.3$\times$10$^{18}$ cm$^{-2}$ (5$\sigma$). The black contours begin at 5.6$\times$10$^{19}$ cm$^{-2}$ in steps of 1.3$\times$10$^{20}$ cm$^{-2}$. The integrated-velocity range is from 105 $\kms$ to 395  $\kms$. The beam of FAST is shown in the bottom-left corner.}
\label{Fig:M101_M_HI-DSS}
\end{figure}

\section{Results}
\label{sect:results}
Figure \ref{Fig:M101_M_HI-DSS} shows a \HI column-density map superimposed on the Digitized Sky Survey (DSS) $B$-band optical image for the M101 galaxy group. The \HI map with an angular resolution of 2.9$^{\prime}$ was construced from the FAST observation.  From Figure \ref{Fig:M101_M_HI-DSS}, we see that the \HI disk around the spiral galaxy M101 much more extends than the bright optical counterpart. The outskirts of the \HI disk are asymmetric, and extend toward the southwest. It has been as also seen in previous single-dish observations and array synthesis observations. Howerver, compared with the previous array synthesis observations \citep{van1988,Walter2008}, our new observations show a more extended structure around M101. The extra extended region is roughly considered as that between blue contour and first black contour in Figure \ref{Fig:M101_M_HI-DSS}. While compared with the early single-dish observations \citep{Huchtmeier1979,Mihos2012}, we find that the extended structure has a tail toward the south. In addition, a dwarf companion NGC 5477 and a massive giant \HII region NGC 5471 reside near the northeast of  the extended disk. The other companion NGC 5474 is located at the south of the extended disk, whose \HI disk also displays an extended structure toward the northwest.  In a \HI column-density map, \citet{Huchtmeier1979} detected a diffuse-gas bridge connecting M101 with NGC 5474, yet our highly sensitive observation with a relatively high angular resolution don't detect the bridge between the two galaxies, as shown in Figure \ref{Fig:M101_M_HI-DSS}.

\begin{figure}
\centering
\includegraphics[width = 0.47 \textwidth]{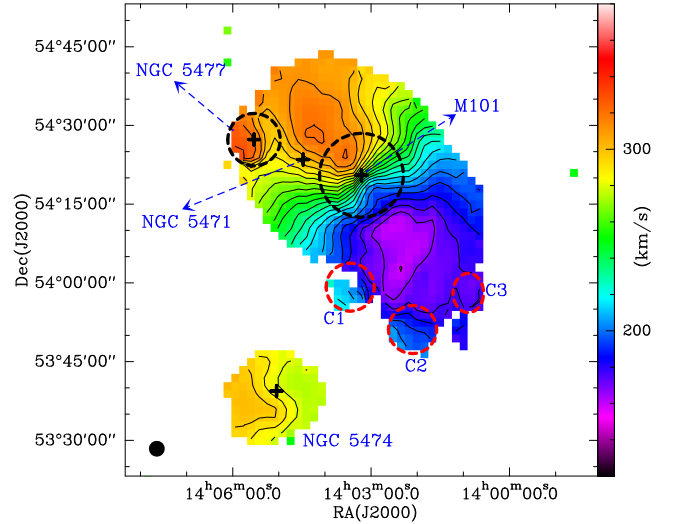}
\vspace{-6mm}
\caption{Velocity field obtained from the \HI data cube (the first moment map), which is shown in both contours and colourscale. The velocity contours go from 120 to 380 $\kms$ in steps of 8 $\kms$. The red dashed circles and  ellipse represent the three newly identified \HI clouds. Letters C1, C2, and C3 illustrate the locations of the three \HI clouds. The black pluses mark the center positions of M101 and members. The NGC 5477 region and the $R_{25}$ (8$^{\prime}$) of the M101's optical disk are shown in two black dashed circles, respectively. }
\label{Fig:M101_velocity}
\end{figure}

\begin{figure*}
\centering
\includegraphics[width = 0.9\textwidth]{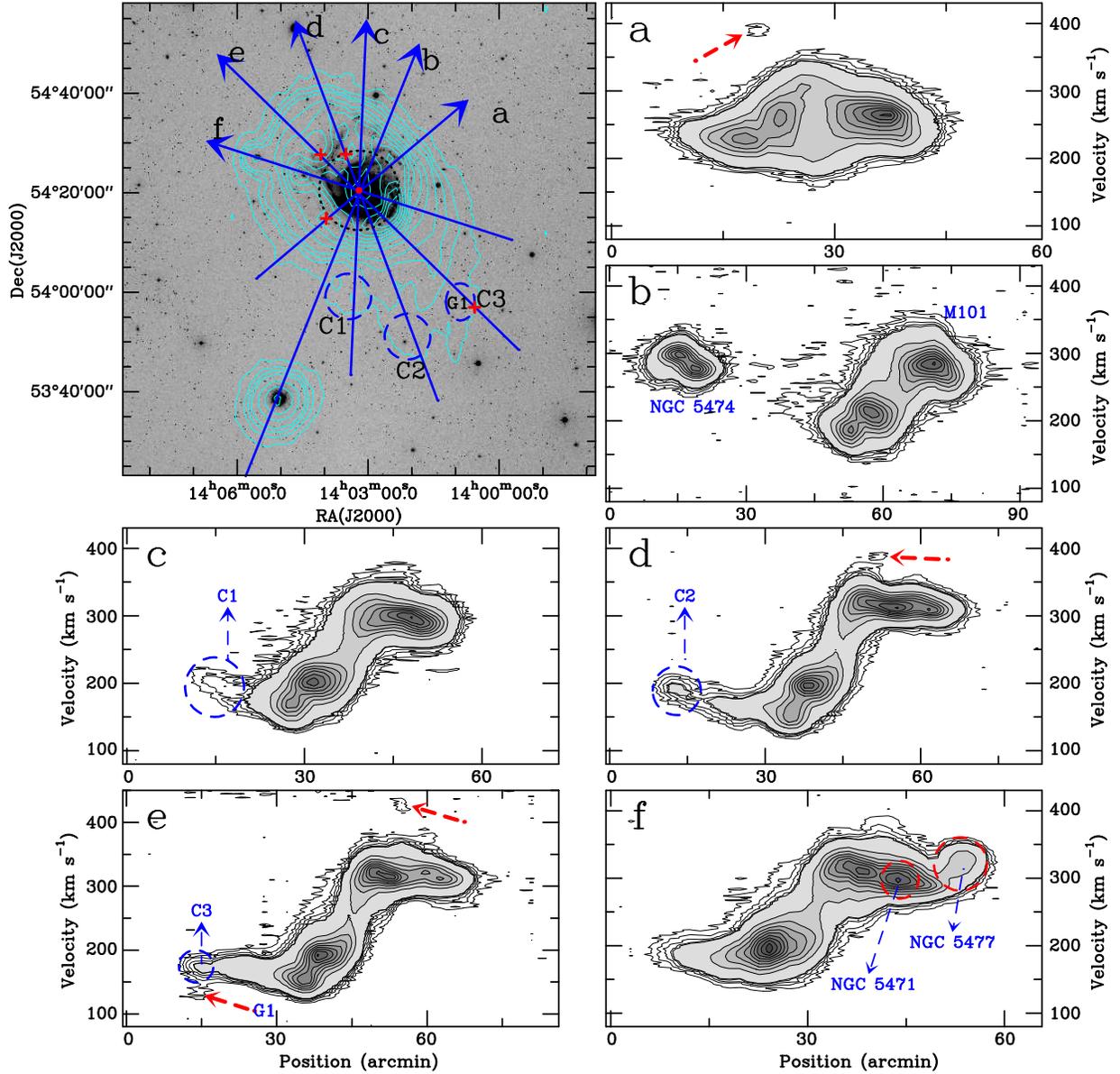}
\vspace{-1mm}
\caption{Six position-velocity plots along different directions. The top-left panel shows \HI column-density contours in cyan colour, overlaid on the DSS  $B$-band optical image. The contour values are the same as those in Figure \ref{Fig:M101_M_HI-DSS}.  The six black arrows mark the directions and positions of position-velocity diagrams in panels a-f, respectively.  The four red pluses mark the four high-velocity \HI clouds. Letter G1 indicates one of the high-velocity \HI clouds. The black dot circle represents the iosphotal size ($R_{25}$) of the M101's optical disk. In panels a, c, d, e, and f, the black contours begin at 2$\sigma$ (112 mK) in steps of 2$\sigma$,  and in steps of 20$\sigma$ for the denser region, while  the black contours begin at 1$\sigma$ (56 mK) in panel b.}
\label{Fig:M101-PV}
\end{figure*}

\begin{figure*}
\centering
\includegraphics[width = 0.32 \textwidth]{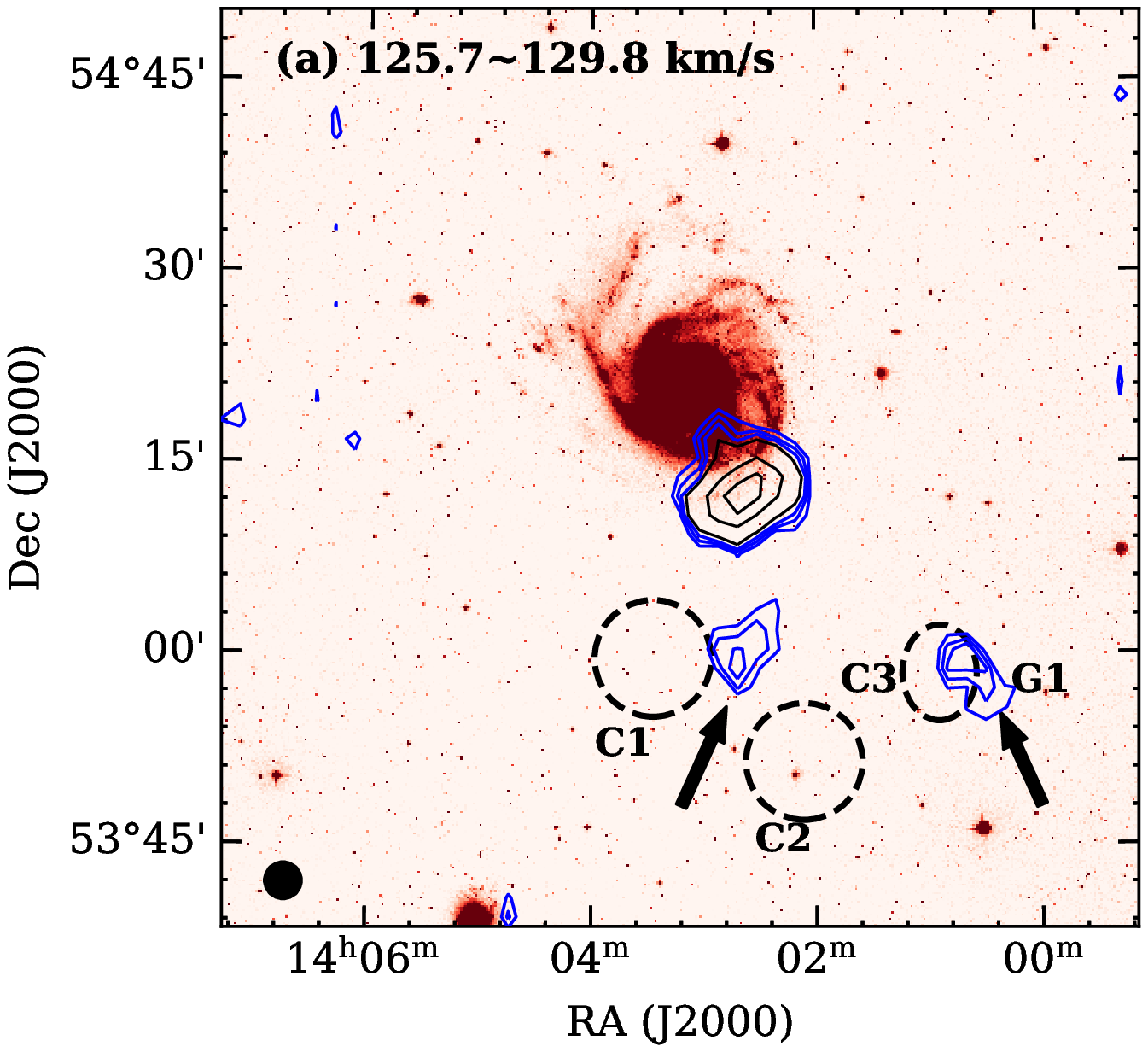}
\includegraphics[width = 0.32 \textwidth]{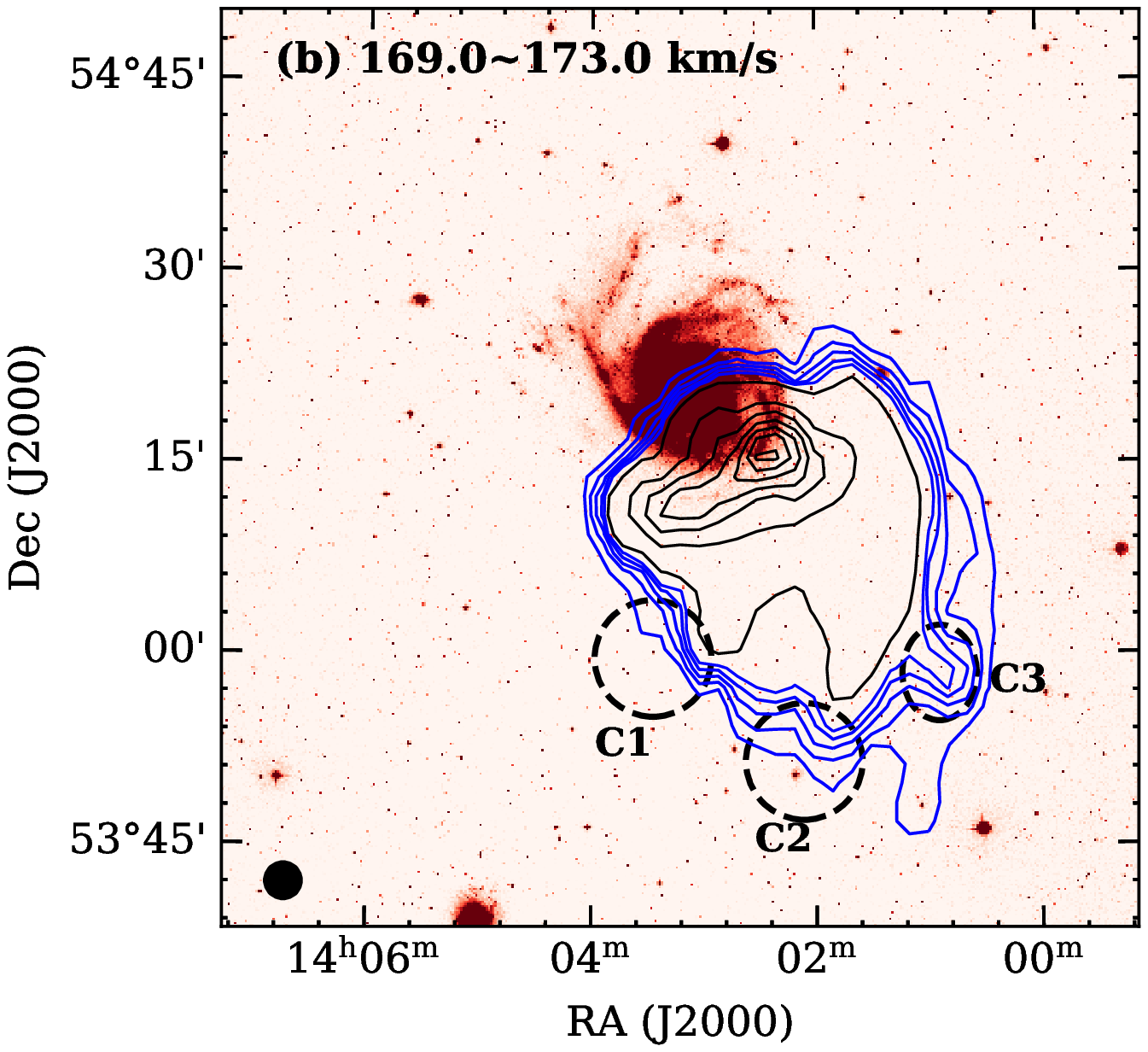}
\includegraphics[width = 0.32 \textwidth]{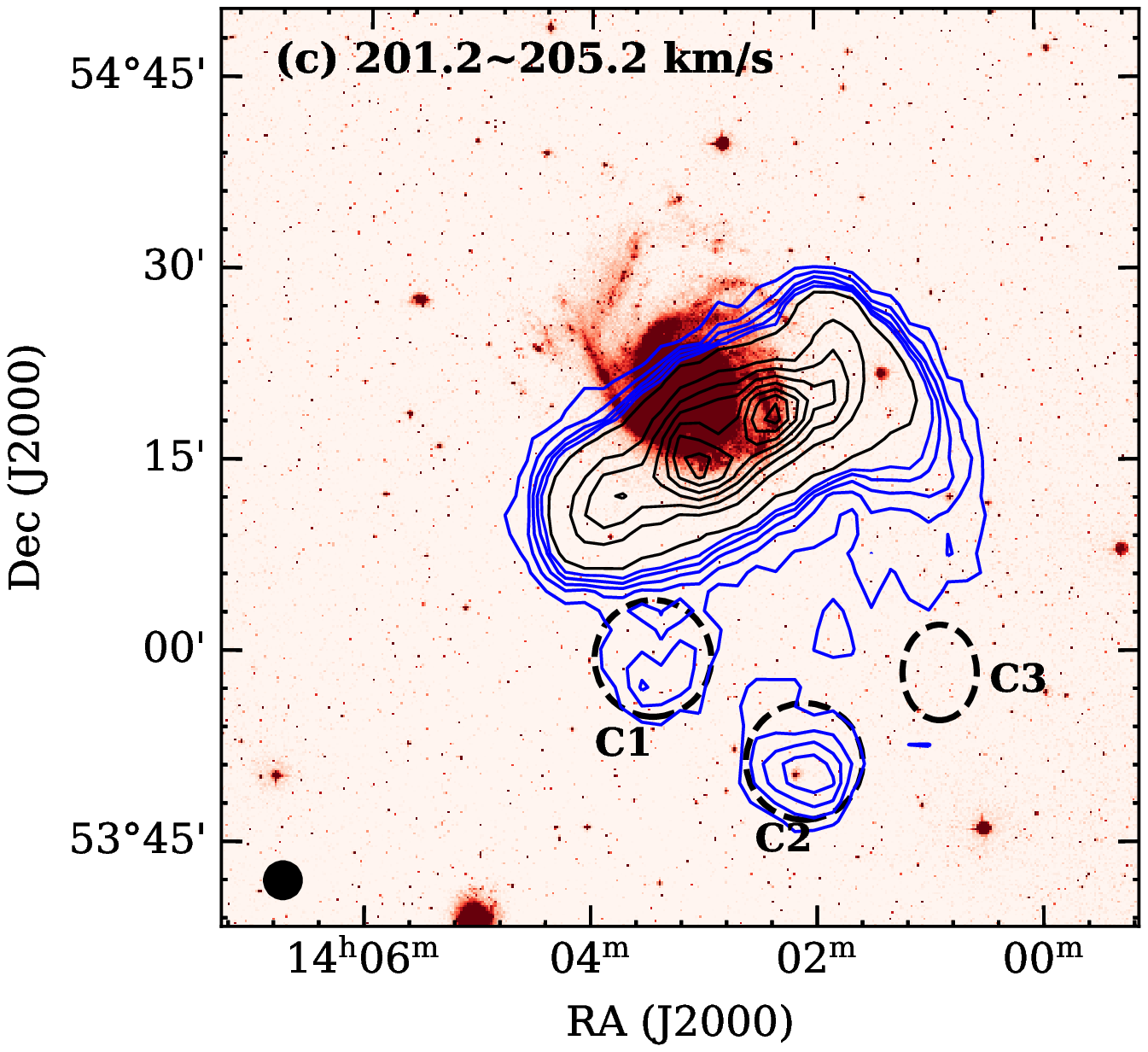}
\vspace{-2mm}
\caption{Channel maps (blue and black contours) overlaid on the DSS  $B$-band optical image. The channel separation and width are about 4 $\kms$. The velocity ranges are shown in the top-left corner of each panel. (a) The two arrows  mark two \HI clouds identified by \citet{Mihos2012}. The blue contours begin at 0.33 K km s$^{-1}$ in steps of 0.20 K km s$^{-1}$,  and in steps of 2.27 K km s$^{-1}$ for the denser region indicated by black contours. (b) The blue contours begin at 0.50 K km s$^{-1}$ in steps of 0.67 K km s$^{-1}$,  and in steps of 13.36 K km s$^{-1}$ for the denser region indicated by black contours. (c) The blue contours begin at 0.50 K km s$^{-1}$ in steps of 0.67 K km s$^{-1}$,  and in steps of 13.36 K km s$^{-1}$ for the denser region (black contours). The dashed circles and ellipse represent the three newly identified \HI clouds. The beam size of FAST is indicated in the bottom-left corner of each panel. }
\label{Fig:M101_M_HI-DSS-H4}
\end{figure*}

Figure \ref{Fig:M101_velocity} shows the \HI velocity field of the M101 galaxy group. The iosphotal size ($R_{25}$) of the M101's optical disk is 8$^{\prime}$ \citep{Mihos2013}.  As seen in Figure \ref{Fig:M101_velocity}, the \HI velocity field of M101 within the optical disk region is regular and looks unperturbed. Outside the optical disk region, the velocity field becomes irregular. While the velocity field of  NGC 5477 shows an irregular velocity gradient, and whose recessing velocity is very close to that of the perturbed northeast region of the \HI disk \citep{Combes1991}, indicating that the velocity field of  NGC 5477 has been combined with that of M101. Furthermore, the \HI velocity field of NGC 5474 is symmetric, and shows a S-like shape. \citet{Rownd1994} indicted that the gas disk of NGC 5474 exists a pronounced warp developing beyond the optical disk. The most exciting thing is that we identify three new \HI clouds located on the southern edge of the M101's \HI disk. The identified three \HI clouds have no optical counterparts. To further analysis and discussion, we name them at C1, C2, and C3, respectively. Especially the C2, whose velocity field are very similar to that of the companion NGC 5477.

In order to explore the dynamic relationship of the companions and \HI clouds with M101, we made position-velocity (PV) diagrams along different directions, which are shown in Figure \ref{Fig:M101-PV}.  From these PV diagrams, we first identify four high-velocity \HI clouds, as indicated by red arrows in panels a, d, and e. Here the gas leaving from the \HI disk is considered as high-velocity cloud (HVC). The positions of the four HVCs are marked in four red pluses in the top-left panel of Figure \ref{Fig:M101-PV}. From the top-left panel, we see that the four HVCs are mainly located outside the optical disk region. One of the HVCs in panel e of Figure \ref{Fig:M101-PV} is indicated by G1.  By comparing with the previously detected high-velocity clouds in M101, we find that the remaining three HVCs are associated with those detected by \citet{van1988}. In panels c-e, we also see that the newly detected \HI clouds C1, C2, and C3 similar to the NGC 5477 in panel f rotate with the \HI gas disk of M101. Additionally, compared with the identified HVCs, the  C1, C2, and C3  have the low relative velocity, and their sizes are larger than those of the HVCs, suggesting that these detected clouds may be intermediate velocity cloud (IVC). Furthermore, at an intensity of 3 mJy beam$^{-1}$ ($\sim$1$\sigma$), \citet{Mihos2012} detected a weak \HI gas bridge connecting M101 with NGC 5474 from their PV diagram. However, in panel b of Figure \ref{Fig:M101-PV}, we don't identify the connecting bridge with nearly the same sensitivity.

\section{Discussion and CONCLUSIONS}
\subsection{New {\rm \HI} clouds detected around M101}
Compared with the previous \HI observations toward the M101 galaxy group \citep{Huchtmeier1979,Walter2008,van1988}, we detected a more extended \HI disk around M101 using FAST with a high sensitivity \HI observation. On the southern edge of the extended \HI  disk, we identify three new \HI clouds (C1, C2, and C3). The velocity field of the \HI cloud C2 very similar to the dwarf companion NGC 5477 shows an irregular velocity gradient. From the position-velocity diagrams in panels c-e of Figure \ref{Fig:M101-PV}, we also see that the \HI clouds C1, C2, and C3 similar to NGC 5477 rotate with the \HI gas disk of M101. Unlike the companion NGC 5477, however, they have no optical counterparts. This situation is also very similar to nearby galaxy NGC 2403.  In the NGC 2403, a newly detected cloud rotates with the main \HI disk, see Figure 6 of \citet{deBlok2014}. 

From the observation of GBT,  \citet{Mihos2012} also found two new \HI clouds on the southwest of the M101's \HI disk. To distinguish the previously identified \HI clouds, we made channel maps overlaid on the DSS $B$-band optical image, see Figure \ref{Fig:M101_M_HI-DSS-H4}. From the panel (a) of Figure \ref{Fig:M101_M_HI-DSS-H4}, we see that except for the \HI gas associated with the main disk, there are two \HI clouds located on the southwest, indicated by two black arrows. By comparing with velocity range and position, the two \HI clouds in panel a should be those identified by \citet{Mihos2012}.  Hence, here the \HI clouds C1, C2, and C3 in panels (b) and (c) of Figure \ref{Fig:M101_M_HI-DSS-H4} are identified for the first time from the M101's \HI disk.

The total mass of the three \HI clouds in M101 can be calculated by $M=2.72mD^{2}\int N(x,y)dxdy$, where the factor 2.72 is the mean atomic weight,  $m$ is the mass of atomic hydrogen, $D$ is the adopted distance  to the clouds, $dx$ and $dy$ are the pixel sizes. The column density $N(x,y)$ in each pixel can be estimated as $N(x,y)= 1.82\times10^{18}\int T_{\rm B}dv$, where $T_{\rm B}$ is the line brightness temperature, and $dv$ is the channel width. We adopt the distance (6.9 Mpc) of M101 as those of the three \HI clouds, yielding that the masses of C1, C2, and C3 are  1.3$\times$10$^{7}$ \msol, 2.4$\times$10$^{7}$ \msol, and 2.0$\times$10$^{7}$ \msol, respectively. The obtained masses are a lower limit because we can only identify cloud emission when they are not projected against the main disk. Meanwhile, we also estimated that the sizes of C1, C2, and C3 are about 18 kpc, 19 kpc, and 15 kpc, respectively. From the Figure \ref{Fig:M101_M_HI-DSS-H4}, we see that the \HI clouds C1  and C3 have connected in projection with the main disk, while the \HI cloud C2 seem not to overlap with the main disk in panel (c). These extra-planar gas in M101 are similar to the filaments found in NGC 891 \citep{Oosterloo2007}, NGC 2403 \citep{Fraternali2002, deBlok2014}, and M33 \citep{Sancisi2008}. \citet{Sancisi2008} suggested that such extra-planar gases are the most remarkable found in the halo regions of these galaxies because they may be direct evidence of cold gas accretion from outside. 

\subsection{The origin of the extra-planar gas in M101}
\begin{figure}
\centering
\includegraphics[width = 0.47 \textwidth]{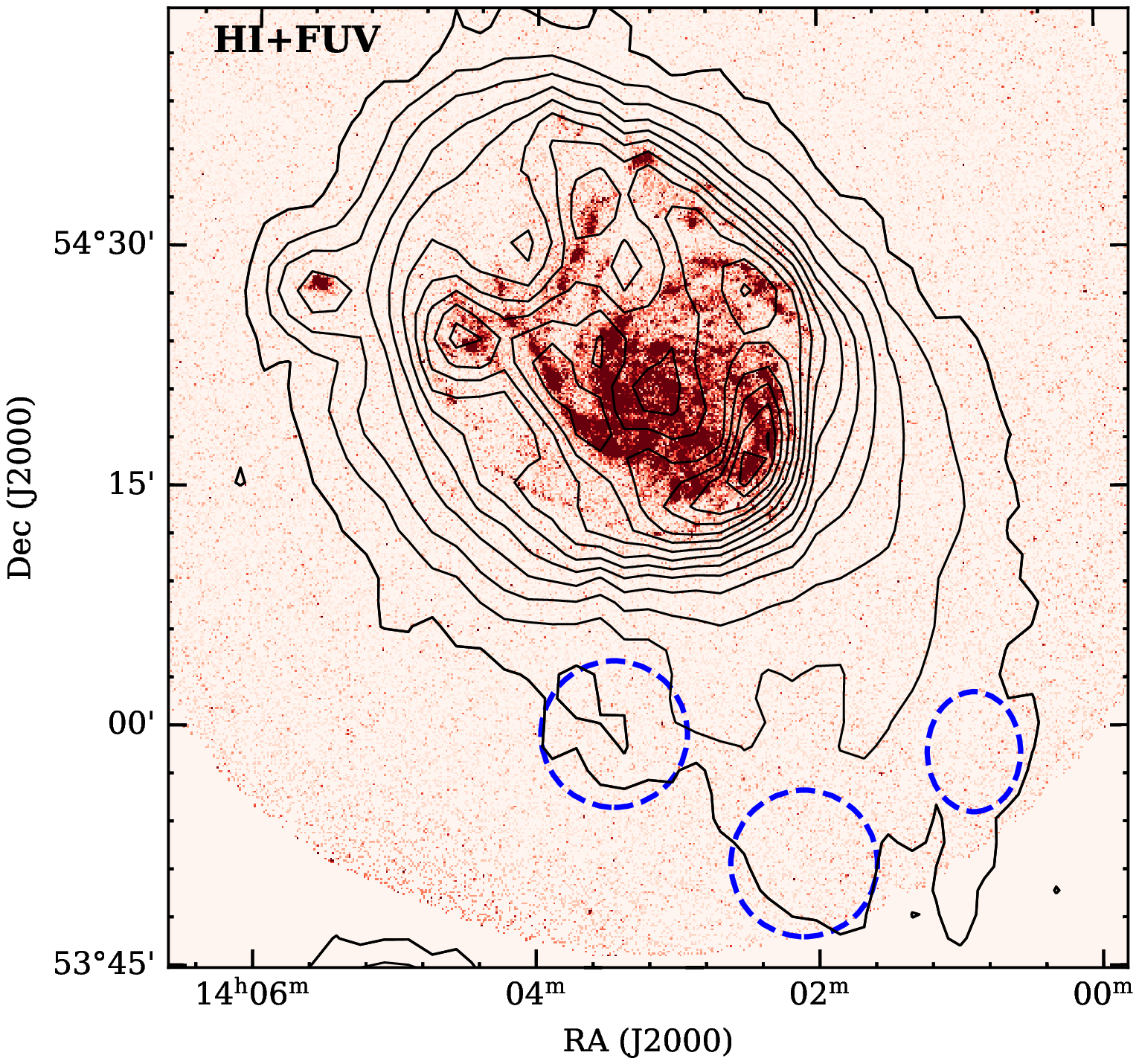}
\vspace{-7mm}
\caption{\HI column-density contours in blue black colours, overlaid on the FUV emission image of M101. The contour levels are the same as those in Figure \ref{Fig:M101_M_HI-DSS}. The blue dashed circles and ellipse represent the three newly identified \HI clouds.}
\label{Fig:M101_HI-FUV}
\end{figure}

For the origin of the extra-planar gas, it seems to has two mechanisms \citep{Sancisi2008}, fountain-driven accretion and accretion from intergalactic space. In fountain-driven accretion, the hot gas rises into the halo, condenses into cold clouds and returns to the disk. Here the extra-planar gas may be driven  by star formation. Figure \ref{Fig:M101_HI-FUV} shows the far-ultraviolet (FUV) emission image of M101 overlaid with the \HI column-density contours. The FUV data, obtained from \citet{Dale2009}, are used to trace emission from hot stars \citep{deBlok2014}.  However, from the Figure  \ref{Fig:M101_M_HI-DSS} and \ref{Fig:M101_HI-FUV}, we did not see the evidences of violent star formation in the projection directions of these clouds and the HVC G1 identified by us. Thus, the fountain-driven accretion can not cause these IVC \HI clouds and the HVC G1 detected in M101. 

While for the other mechanism, the extra-planar gas is either directly accreting from the intergalactic medium or is the result of a minor interaction with a neigboring dwarf galaxy, as the case in NGC 2403 \citep{deBlok2014}. According to \citet{Dekel2006} and \citet{Mihos2013}, if massive galaxies at low redshift undergone a cold flow accretion, they will host a massive hot halo. While the hot gas traced by the X-ray emission in M101 is diffuse and only correlated with the spiral arms \citep{Kuntz2003}. Furthermore, as shown in Figure \ref{Fig:M101_HI-FUV}, the FUV emisson in M101 is also similar to the X-ray emission. Thus, the M101 does not host a massive hot, and even warm halo, arguing that the massive galaxy M101 ($M_{\rm tot}\simeq 10^{12} \Msun$, \citet{Huchtmeier1979}) would not be experiencing any significant cold flow accretion from the intergalactic medium at the present day \citep{Mihos2013}. Simultaneously, it also means that the main galaxy M101 may not have enough energy to eject such massive clouds found by us and \citet{Mihos2012}. Hence, the extra-planar gas in M101 may be the result of a minor interaction with a neigboring dwarf galaxy, but there is still a lack of reliable evidence for the interaction.

From Figure \ref{Fig:M101_velocity}, we see that the \HI velocity field  within the optical disk region of M101 is regular, indicating that the relatively strong disturbance occurs in the outer disk.  While the asymmetric outer disk of M101 has also been believed to arise from an interaction with a companion \citep{Beale1969,Rownd1994,Waller1997,Mihos2013,Mihos2018}. The two closest companions to M101 are NGC 5477 and NGC 5474.  NGC 5477 has a low luminosity ($M_{V}$ =-15.3).  \citet{Mihos2012} suggested that it can't be massive enough to creat such a strong morphological response in a giant Sc spiral like M101. For the other companion NGC 5474,  it is more luminous than NGC 5477.  \citet{Huchtmeier1979} and \citet{Mihos2012} detected some diffuse \HI gas between M101 and NGC 5474 in a \HI  column-density map and a PV diagram, respectively, strengthening the case for a recent interaction between the two galaxies. 

While our high sensitivity \HI observation with a relatively high resolution (2.9$^{\prime}$) did not detect the diffuse \HI bridge connecting M101 with NGC 5474 in a \HI  column-density map (see Figure \ref{Fig:M101_M_HI-DSS}) in a PV diagram (see the panel b of Figure \ref{Fig:M101-PV}). From Figure \ref{Fig:M101_M_HI-DSS}, we obtain that the distance between M101 and NGC 5474 is about 15$^{\prime}$. While the angular resolutions of the previous obsevations from \citet{Huchtmeier1979} and \citet{Mihos2012} are about 9$^{\prime}$, which is nearly close to the distance between the two galaxies. Hence, above two observations may not be able to distinguish the gas between them well. We suggest that the previously detected bridge between the two galaxies is likely due to beam dilution created by the role of low spatial resolution. Generally, galaxy-galaxy interaction can give rise to star formation in tidal debris, while \citet{Garner2021} did not detect an abundance of star formation between  M101 and NGC 5474, also suggesting that the gas between the two galaxies is very weak or there is no gas like bridge. We did not detect a \HI bridge between the two galaxies, but identify a slightly extended \HI disk for NGC 5474. It means that the two galaxies have undergone only a weak interaction, just like the judgement of \citet{Garner2021}.

In addition, as shown in Figure \ref{Fig:M101_M_HI-DSS}, we detected a new \HI tail on the southern edge of the extended \HI disk of M101. Generally, the \HI tail indicates ongoing minor merger and recent arrive of external gas, and then can be considered as direct evidence of cold gas accretion \citep{Sancisi2008}. Compared with the positions of the \HI tail, the detected new \HI clouds seems to can represent other \HI tails.  Here the \HI tails detected in M101 gives a reliable evidence for M101 interaction with NGC 5474. Thus, we conclude that the extra-planar gas (three \HI clouds) and the \HI tail detected in the M101's disk are likely to origin from a minor interaction with the companion NGC 5474. Furthermore, because FAST has high sensitivity and relatively high angular resolution, it can detect and separate the relatively weak and small gas structures around galaxies. This will play an important role in the study of galaxy evolution. 

\acknowledgments 
We thank the referee for insightful comments that improved the clarity of this manuscript. We acknowledge the supports of the National Key R$\&$D Program of China No. 2018YFE0202900. 
This work was also supported by the Youth Innovation Promotion Association of CAS, the National Natural Science Foundation of China (Grant No. 11933011), and supported by the Open Project Program of the Key Laboratory of FAST, NAOC, Chinese Academy of Sciences.

\end{document}